# Biological Microscopy with Undetected Photons

**András Búzás[1], Elmar K. Wolff[2], Mihály G. Benedict[3], Pál Ormos[1] and András Dér[1]**
[1]Institute of Biophysics, Biological Research Centre of Szeged, Temesvári krt. 62, H-6726 Szeged, Hungary
[2]Institute for Applied Biotechnology and System Analysis at the University of Witten/Herdecke, Stockumer Str. 10, 58453 Witten, Germany
[3]Department of Theoretical Physics, University of Szeged, H-6720 Szeged, Tisza Lajos krt 84-86, Hungary

Corresponding author: András Dér (e-mail: der.andras@brc.mta.hu).

This work was sponsored by a Hungarian research grant, NKFI-6, K-124932, and a collaboration grant of the DFG, WO 1713/2-1.

**ABSTRACT** Novel imaging techniques utilizing nondegenerate, correlated photon pairs sparked intense interest during the last couple of years among scientists of the quantum optics community and beyond. It is a key property of such "ghost imaging" or "quantum interference" methods that they use those photons of the correlated pairs for imaging that never interacted with the sample, allowing detection in a spectral range different from that of the illumination of the object. Extensive applications of these techniques in spectroscopy and microscopy are envisioned, however, their limited spatial resolution to date has not yet supported real-life microscopic investigations of tiny biological objects. Here we report a modification of the method based on quantum interference by using a seeding laser and confocal scanning, that allows the improvement of the resolution of imaging with undetected photons by more than an order of magnitude, and we also present examples of application in the microscopy of biological samples.

## I. INTRODUCTION

During the recent years, an exciting new paradigm of quantum imaging has emerged [1-5], with possible implications in various branches of microscopy, spectroscopy and information technology. Related measuring techniques allow observation (imaging) at wavelength ranges different from those of the absorption of targeted objects ("imaging with undetected photons" [1,4], referred to as "UP-imaging" in the following) offering serious advantages, e.g., in infrared imaging important in medical, industrial or forensic applications (e.g., cancer diagnostics) [6]. Although recent works have shown that imaging with undetected photons is possible also by "classical" light [7], [8], original works utilizing quantum mechanical principles remained in the focus of interest [9], [10]. These methods are based on the generation of nondegenerate, correlated (entangled) photon pairs of different wavelengths in nonlinear media, by the method of optical parametric down-conversion. While quantum entanglement and other quantum mechanical effects [11]-[13] are a commonplace in the world of subatomic particles, they usually remain hidden in macroscopic phenomena. Nevertheless, they play a vital role in such important disciplines of the „macro"-world science as laser physics, quantum computing, or the emerging field of quantum biology, as well [14]. A recent, comprehensive review of nonlinear optics and spectroscopy with quantum light has been presented in [15]. Among the various alternatives of implementation of quantum imaging, perhaps the most promising ones are using the quantum interference (QI) approach [1,4]. It has numerous practical advantages over alternative solutions, which require either the simultaneous, synchronized detection of both down-converted photons of different frequencies (e.g., in the case of quantum ghost imaging [2]), or high coherent light intensities (e.g., in the methods utilizing optical parametric up-conversion [5] or optical parametric amplification (OPA) [3]), thereby increasing technical complexity, or hampering the imaging of fragile samples, such as biological ones [4].

Application of quantum imaging methods in the microscopy of biological or other samples having tiny feature sizes, on the other hand, requires sufficient spatial resolution. Typical resolutions in cellular imaging are supposed to be in the range of a few microns, or better. Conventional microscopy goes down to the diffraction limit determined by the wavelength of the observed light, while recent super-resolution techniques improve it up to an order of magnitude [16]. In some histopathological imaging techniques, on the other hand, infrared monitoring light is used in order to reduce background scattering. While here a resolution in the range of microns is usually considered sufficient [17], as a recent development in deep-tissue imaging, a novel three-photon microscopy technique allowed a submicron resolution imaging of a mouse brain across the skull [18].

The spatial resolution of quantum ghost imaging, however, is inherently limited by the strength of the correlation between the entangled photon pairs, determined by the properties of the nonlinear crystal and the pump beam, in addition to the point spread function associated to other parts of the optics and the detector, as Padgett et al. have recently revealed [19]. In a microscopy technique based on ghost imaging, they recorded a raw image of a wasp wing, and established a resolution of about 15 µm [20], [21]. Although the spatial resolution of QI methods presented by Zeilinger et al. is not specified in their publications [1, 4], it must also be limited by the numerical aperture (NA) of the imaging objective adjusted to other characteristic features of their setup, such as the 4f-arrangement of their optics, or the size of the nonlinear crystal used for the generation of entangled photon pairs. From the images presented in [1] and [4], we estimate their resolution to be in the range of 20-30 micrometers, which also does not offer extensive applications in biological microscopy. Nevertheless, in their patent publication [4], Zeilinger et al. show an image and a video of a zebra fish embryo of a sub-mm feature size. To our knowledge, examples in [4], [20] and [21] are the first ones published on quantum imaging of biological samples.

Here we describe a technique utilizing the interference of photons created by parametric down-conversion, similarly to the approach of Zeilinger et al. [1, 4], but with a considerably improved spatial resolution, readily allowing applications in the infrared microscopy of biological (and other) samples. After a technical description of the method, examples of typical images are presented below, and further advantages and limitations of the new approach are discussed.

## II. RESULTS

The main goal of our work was to improve the spatial resolution of the UP-imaging technique, in order to make it applicable in biological microscopy. In classical diffraction-limited optical imaging, the primary factors that determine the resolution limit ($d$) are the numerical aperture (NA) of the objective lens collecting the light coming from the sample, and the wavelength ($\lambda$): $d = 0.61\lambda$ / NA (Rayleigh criterion). In the practical implementations realizing the concept of QI [1,4], the setups were optimized for the image quality, using high-efficiency production of entangled photon pairs with relatively sizable ppKTP crystals: 1 by 2 by 2 mm$^3$, each. Along their optical axis, the crystals contain periodically poled rectangular sheets of 9.325 µm period. In turn, this geometry restricts the NA of the lenses focusing the exciting light into the nonlinear crystals, wherein quasi-planar wavefronts are required to obey the conditions for down-conversion. In the beam waist of focused Gaussian beams of low angular spreads ($\Theta$), this holds approximately for twice of the Rayleigh distance ($2z_R = 2w_0/\text{NA}$, where $w_0$ is the waist radius), which should not be smaller than the crystal length ($l_c$) (SI Appendix, Figure S1). In fact, $2z_R = l_c$ (= 2 mm) was realized in [1] and [4], maximizing the NA of the focusing length and the imaging system in general, due to the 4f arrangement used in these experiments. According to the Rayleigh criterion, the resolution limit is maximal under these conditions, and, using the estimated divergence value of 50 mrad for the idler photons [22], it is calculated to be about 16 µm, which is close to the estimated resolution of the images published in [1] and [4]. (Note that the wavelength conversion provided by the setup does not affect the resolution.) Based on the above arguments, one might assume that the resolution limit could be improved by choosing a thinner NLO crystal with stronger focusing, sacrificing brightness to gain resolution. However, concerning that the curvature of wavefronts at $z_R$ grows nonlinearly with focusing (the radius of curvature, $R(z_R) = 2\pi w_0^2 / \lambda$), an additional restriction also holds for nonlinear optical crystals used in parametric down-conversion, namely, that the non-correlated background drastically increases if the angular spread of the exciting light exceeds a certain limit (established to be ca. 32 mrad in [23]), that is only slightly higher than the one used in [1] and [4] (ca. 26 mrad), not allowing a considerable improvement in resolution.

To break this apparent limit, our concept was to introduce a confocally arranged pair of objectives (i.e., high-NA lenses) into the collimated path of measuring light (O1 and O2 in Fig. 1) to illuminate the sample (S) in the common focal plane, and gathering the light passing through it. According to the Rayleigh-criterion, this arrangement allows a much higher resolution of the sample scanned by the imaging light, depending on the NA of O1 and O2 ($d \approx 2$ µm for NA = 0.4). Note that the NA value of the objectives is not limited in this arrangement, *per se*, so they can be chosen arbitrarily, allowing a high spatial resolution. Taking into account the thickness of biological samples, however, we chose an intermediate value of 0.4 for our experiments, in order to maintain a decent field depth, as well.

The use of confocal illumination and observation is similar to that in confocal scanning microscopies (CSM) [24]. Unlike in most CSM techniques, however, our method does not apply a pinhole and does not require a fluorescent sample either, but relies on the observation of optical interference between the reference and the sample beams (see below). For the sake of simplicity, we refer to our type of method that combines the UP-imaging features with point-by-point scanning, by the acronym SIMUP, standing for Scanning Interference Microscopy with Undetected Photons. Note that scanning-assisted imaging has been applied in quantum imaging applications [25], too, to break the Rayleigh limit of conventional imaging. We had a less ambitious goal, namely, to improve the lateral resolution of UP-imaging, in order to demonstrate its applicability in biological microscopy. The schematic layout of the measuring setup is shown in Figure 1.

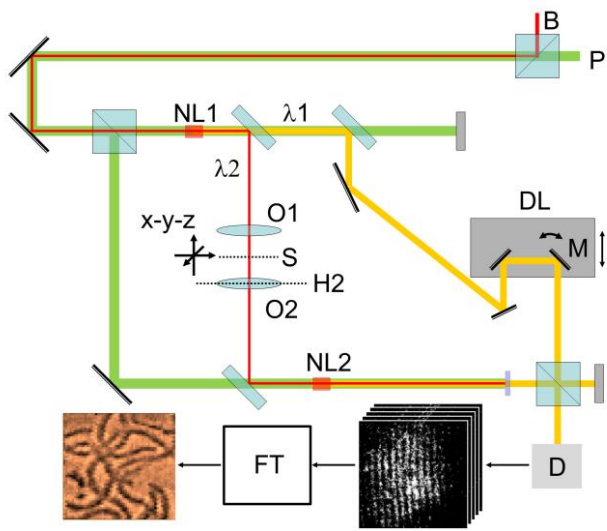

**Figure 1.** Schematic representation of the experimental setup. The exciting green laser light (P, 532 nm), and the entangled daughter beams (orange: 880 nm, red: 1345 nm) induced in the nonlinear crystals (NL1 and NL2) by optical parametric down-conversion. According to the convention, we refer to the shorter-wavelength (880 nm) beam as "signal", and the longer-wavelength (1345 nm) one as "idler". The signal and idler beams are deflected, separated and reunified before the camera (D) by the corresponding normal or dichroic mirrors, beam splitters and filters. A pair of confocally arranged microscope objectives (O1 and O2) was focusing the idler beam (1345 nm) onto the sample (S), which was moved by a complex X-Y-Z scanner stage prior to and during the measurements. The principal plane of O2 is denoted by H2. A precision delay unit (DL) equipped with a rotating mirror (M) was responsible to achieve interference fringes of the 880-nm light on the screen of the camera. The interference images were Fourier-transformed (FT), and the images were reconstructed by a computer. A background laser illumination (B, 1342 nm) was used to enhance conversion efficiency in NL1. (For details, see the Materials and Methods section and the SI appendix.)

Entangled photon pairs are induced by a coherent, visible pump beam, (P, cw Nd-YAG laser, $\lambda$ = 532 nm) in two identical nonlinear crystals (NL1 and NL2, ppKTP), via optical parametric conversion (Type 0, $\lambda_1$ = 880 nm – "signal", and $\lambda_2$ = 1345 nm – "idler").

Idler photons from NL1 pass through the sample, and subsequently enter NL2. Here a very interesting and counterintuitive quantum-phenomenon occurs, the so-called induced coherence by indistinguishability [26]. Namely, if idler photons coming from NL1 are perfectly aligned (and matched in polarization) with those induced in NL2, an observer behind NL2 cannot distinguish the source of these photons (according to its authors this fundamental point of Ref. [26] was suggested by Z. Y. Ou), and merely this fact is enough to induce a second-order coherence of their signal photon counterparts [26], which can be detected as an interference image by the camera. We note that, in a somewhat different arrangement from that of Ref. [26], the first idea of aligning each of the two idlers with pump waves in a parametric down conversion process was proposed in Ref. [27]. Regarding applications to biological samples, the parameters of the crystals are chosen such that the wavelength of the down-converted idler photons hitting the sample ($\lambda_2$) is outside of the main absorption peak of water in the region (centered around 1420 nm), while the wavelength of the detected signal beam ($\lambda_1$) lies in the sensitivity range of the CCD cameras (< 900 nm).

The sample was held by a computer-controlled X-Y-Z stage (mechanical + piezo), that allowed stepwise scanning in the X-Y plane, and fine adjustment of the focus along the Z axis. An interference image was detected by the camera (D) at each position of the sample, and sent to a computer. In order to have interference fringes at the detector plane (Figure 1), mirror M was tilted such that the wavefronts of the signal beams incident on the camera from the two paths (originating from NL1 and NL2, respectively) made an angle of a few degrees, determining the number of stripes per fringe image to be typically 10. From the position and contrast of the fringes, a computer program based on Fourier-transforming the images, assigned an amplitude and phase value to each position of the sample, from which amplitude and phase images of the object were reconstructed. If, e.g., the optical pathlength through the sample changed from one point to another during scanning, the interference fringes also shifted accordingly, to the "left" or the "right", depending on the sign of the change. If, however, the transmission of the sample increased or decreased, the contrast of the fringes followed this change, respectively. (For details, see the theoretical treatment below, and Fig.S2 of the SI Appendix.) In order to increase the signal-to-noise ratio of the interference fringes, we applied an additional weak, continuous laser beam (B) of wavelength essentially indentical with that of one of the secondary beams (idler) after the crystals. This beam was also aligned with the identical directions of the idlers in both nonlinear crystals, as shown in Fig. 1. In this way the presence of the seeding laser does not alter the indistinguishability of the photons in the common idler mode. Nevertheless, in this case, the conditions for the interference of signal photons on the camera are ensured also by another effect, namely induced coherence by a laser [27]. In the Appendix, we present the outline of a quantum-optical calculation for the visibility of interference patterns both in the presence and absence of the seeding background (Fig. S5), by generalizing the model of refs. [28], [29]. In the absence of background illumination, when the number of photons originating from the crystals is low, the maximal visibility for the interference of signal photons at the camera should be linearly dependent on the amplitude transmission (*t*) of the sample. This is ideally fulfilled close to the „quantum limit", where one can speak about the interference of photons on the camera, instead of that of classical beams. On the contrary, for large photon numbers when approaching the classical limit, the visibility vs. amplitude transmission curve becomes increasingly nonlinear. Based on our measurements (i.e., the number of photons captured by the camera at controlled transmissivities, the amplification factors at NL1 and NL2) and on our quantum optical calculations (SI Appendix), we determined the normalized photon number values in our experiment with and without background illumination (see details in the SI appendix and in Materials and Methods). The experimental data are satisfatorily maching the theoretical curve, showing only a moderate deviation from linearity (Fig. S5).

Another important consequence of using the seeding laser is due to its relatively long coherence length (ca. 7 mm). Accordingly, the limiting role of the crystal length in the correlation of the signal and idler photons [19] is not dominant here. As long as we can detect the interference of the signal beams by the camera, the resolution is limited primarily by the beam waist in the sample plane, just as in the case of classical confocal imaging. Hence, the fact that the visibility of the interference pattern of the signal photons is a function of the pump waist (see [31], [32]) is also of secondary importance, contrary, e.g., to the case of quantum ghost imaging [19]. (In other words, in the present setup, the existence of an interference pattern is the actual prerequisite of imaging rather than its contrast.) From this point of view, therefore, the resolution of our imaging setup is the same as that of a classical one, namely, determined by the point spread function in the beam waist, ideally given by the Rayleigh criterion.

Figures 2a and b show 2D images of a test object (T), a rectangular grid of regularly spaced, thin photopolymer stripes exposed on the surface of a glass substrate. Although T was a pure phase object, the evaluation method yielded an amplitude image, as well (Fig. 2a). Note, however, that the reconstructed phase image of T (Fig. 2b) has a spatial resolution better than that of the amplitude image (Figures 2c and d).

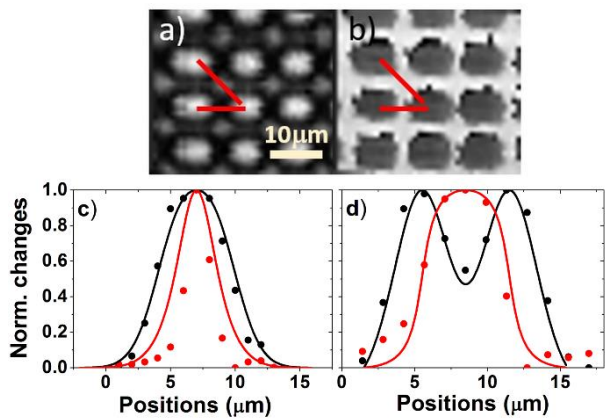

**Figure 2.** Amplitude (a) and phase (b) images of the test object, T (a rectangular grating of transparent photoresist stripes). Normalized intensity distributions of the images were determined along cross sections represented by the red lines of a) and b). The results are depicted by the filled circle symbols in c) for the shorter and d) for the longer lines, respectively. The solid lines represent the results of wave-optic simulations. Black lines and symbols stand for the amplitude, while the red ones are for the phase images. The best fit to the measured data were obtained by line widths of 2 μm and 6 μm (c and d, respectively), and beam waist $w_0 = 3$ μm. The phase difference value was taken from the experiments: $\Delta\phi = 3.7$.

In order to reveal the origins of the amplitude and phase images, and to understand the difference between their resolution, we carried out model calculations mimicking the imaging conditions (Fig. 3). For this purpose, a simplistic approach was used to compute the average amplitude and phase of light in a small section around the optical axis in the principal plane of O2 ("lens plane"), resulting from the interference of light waves diffracted from the sample in the common focal plane of O1 and O2 ("focal plane"). Since the rest of the imaging system serves to visualize the intensity and phase in the lens plane (H2 in Fig. 1) via detecting the amplitude and position of interference fringes by the camera, here it is enough to take only this part of the light path into account. The object was considered planar, while the intensity and phase conditions at H2 were calculated using the Rayleigh-Sommerfeld approximation of the scalar diffraction theory [33].

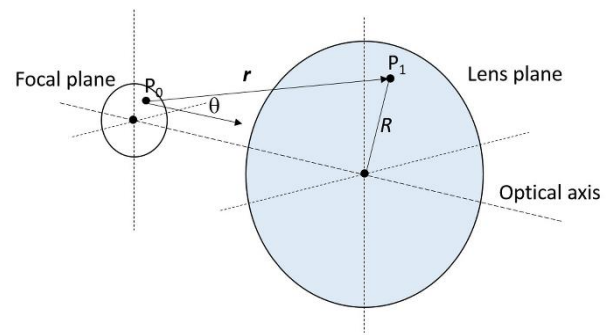

**Figure 3.** Schematic representation of the geometry used for the model calculations. The focal spot is in the idealized sample plane, and the "lens plane" is the principal plane of the O2 objective (H2 in Fig. 1). **r** is a vector directed from P0 to P1, *and R* is the distance between the optical axis and P0. $\Theta$ is the angle the **r** vector makes by the optical axis.

The effect of an amplitude or a phase object located in the focus, S can be described as follows:

$$U(P_1) = -\frac{i}{\lambda} \iint_{FP} U(P_0) \frac{\exp(ikr)}{r} \cos\theta \, ds \quad (1)$$

where $U(P_0)$ is the complex amplitude of the electromagnetic field at a point $P_0$ in the focal plane (FP), $k$ is the wave number, and ds is the surface element. $U(P_1)$ stands for the same at point $P_1$ of the lens plane, $r = |\mathbf{r}|$ where **r** is a vector directed from $P_0$ to $P_1$. The integration runs over the whole aperture of the focal plane. In the focal plane, a Gaussian beam approximation was used with intensity distribution of $w_0 = 3$ μm Gaussian width, and a planar phase front was assumed at the focal plane. The size of the aperture considered around the focal spot was 20 μm (large enough compared to $w_0$, so that the boundary conditions do not influence the results of the calculations). For the calculation of the electromagnetic field at the lens plane, the effect of the lens was taken into account as a phase transformation ($\Delta\phi$) given below:

$$\Delta\phi = 2\pi \frac{\sqrt{R^2 + f^2} - f}{\lambda} \quad (2)$$

where $f$ is the focal length of the lens, and $R$ is the distance between the optical axis and $P_0$. The effect of beam propagation after the lens was neglected.

The light intensity ($I$) in the lens plane was calculated by integration over an aperture of 0.5 mm by 0.5 mm (corresponding to the observation area on the camera), and the phase ($\phi$) was calculated by averaging the phase *ibidem*:

$$I = \iint_{\text{Aperture}} |U|^2 dS \quad (3)$$

$$\phi = \langle \text{phase}(U) \rangle_{\text{Aperture}} \quad (4)$$

The standard deviation of the phase over the integration area was less than 4%. Note that due to the linear nature of the above equations, a homogeneous transmission or phase change introduced by a sample in the focal plane is identically transferred to $I$ and $\phi$, respectively. Due to the phenomenon of induced coherence by indistinguishability, these features of the $\lambda_1$ beam incident into NL2 induce proportional synchronization and coherence in the outgoing signal beam, as it was shown in [26]. In turn, $I$ and $\phi$ can be revealed by detecting the interference of the signal beams by the camera. (For demonstration, see SI Appendix, Figure S2.) Note that the wavelength conversion from idler at the sample to signal at the camera does not alter phase and amplitude information, therefore, given a confocal, point-by-point imaging, it does not alter the lateral resolution of the system, either.

Using the above formulas, numerical calculations were performed in 100-nm steps for scanning areas of 0.5 mm x 0.5 mm, to mimic the results of our experiments. Both for virtual phase and amplitude objects, scanning with stripes of 2 and 6 μm widths in the focal plane (corresponding to the disjoint line width and that at the junctions) were performed, to monitor the effect of feature size. Comparison of the measured and simulated intensity profiles for test phase objects is shown in Figs. 2c and d. The value of phase difference used in the simulations ($\Delta\phi = 3.7$) was adopted from the experiments. The results nicely account for the existence of both amplitude and phase images in the case of pure phase objects, and also for the observed difference between the spatial resolution of the two cases. Note that the simulations also reproduced the appearance of a dip in the case of thicker lines, seen in the amplitude images at the junctions of the photo-polymerized stripes. The reason behind the dip in the "amplitude image" of the phase object is a kind of edge effect. Namely, part of the light diffracted at the border of the phase object does not reach the area of observation on the camera, and this light intensity loss appears to be an extinction in the evaluation.

Fit to the phase image got by scanning with the 2-μm stripe shows higher uncertainty than that to the amplitude image, which is attributed to the higher sensitivity of the phase image to relative inaccuracies of the object width observed by scanning electron microscopy (SI Appendix, Figure S3). The 6-μm line width, on the other hand, is apparently large enough to allow a decent estimate of the edge resolution of the phase image, which is supposed to be determined by the convolution of the transfer function of the imaging system and the phase profile of the object. From model fitting to the measured data (Figure 2d), we claim the edge resolution of the system (distance required for the edge response to rise from 10% to 90%) to be 2 μm, with an estimated uncertainty of about 10% (coming from the inaccuracy of the model fit and the object width). (Since the line spread function ($l(x)$), the one-dimensional extension of the point spread function, is simply the derivative of the edge response ($e(x)$), i.e., $l(x) = d[e(x)] / dx$, the edge resolution is a valid measure of the lateral spatial resolution of a system symmetrical to the optical axis, like ours.) Note that the Rayleigh criterion also gives a 2-μm resolution with the 0.4-aperture imaging objective. Figure S4 of the SI Appendix shows that, according to the model calculations, the spatial resolution of the amplitude image in the case of a pure amplitude object is about 3.8 μm, under similar conditions.

An inference from the above results is that, whenever possible, it is worth recording a phase image by this technique, but the somewhat lower resolution of the amplitude image may still be sufficient to allow the investigation of some tiny objects. Below, we show two examples for the application of SIMUP to image different types of biological samples (Fig. 4).

Spirulina, belonging to the phylum of cyanobacteria, forms helical filaments of typically 50 to 500 μm in length, depending on the actual conditions and strain. At the wavelength of illumination (1345 nm) it can be considered as a mainly phase object, since chlorophyll absorption is negligible in this region, and scattering effects are also considerably reduced in the SWIR regime, as compared to the visible [13], [34], [35]. Figures 4a, b and c, d show amplitude and phase images of a spirulina cell culture, respectively. Note that the phase images have higher contrast, in agreement with the results of the test measurements.

The other example chosen is a wing of a fruit fly. Due to its hierarchically organized structure, it shows characteristic features at different scales: Thick and thinner veins are dominating the mm and 100-μm scale, respectively, while thin, tapered, hair-like formations appear on the micrometer scale. The tip of the hairs is thinner than a micrometer, but the roots are in the range of a few microns. Since all these features show high absorption in the visible and near IR range, the wing represents a principally amplitude object. Figures 4 e, g and f, h show conventional and corresponding SIMUP images, respectively, revealing the structure of the wing at two different scales.

Conventional microscope images were taken in the visible (with an 550-nm filter), hence their resolution is around 300 nm. For comparison of image (g) with the corresponding SIMUP picture (h) obtained with scanning at 1345 nm, a simple image analysis reveals that the circular objects in (g), of a ca. 5-μm diameter (i.e., the "roots of the hairs"), can be transformed to their SIMUP counterparts by blurring the former with a 2-μm wide Gaussian (Fig. S6b), in agreement with the simulation results (Fig. S4, red line). It can be established that the resolution of the SIMUP images exceeds by an order of magnitude or more the resolution of pictures published so far with other methods utilizing quantum imaging with undetected photons [1], [4], [36]. For the biological objects we studied, except for the out-of-plane sub-micron structures, all characteristic features are faithfully reflected by the SIMUP images.

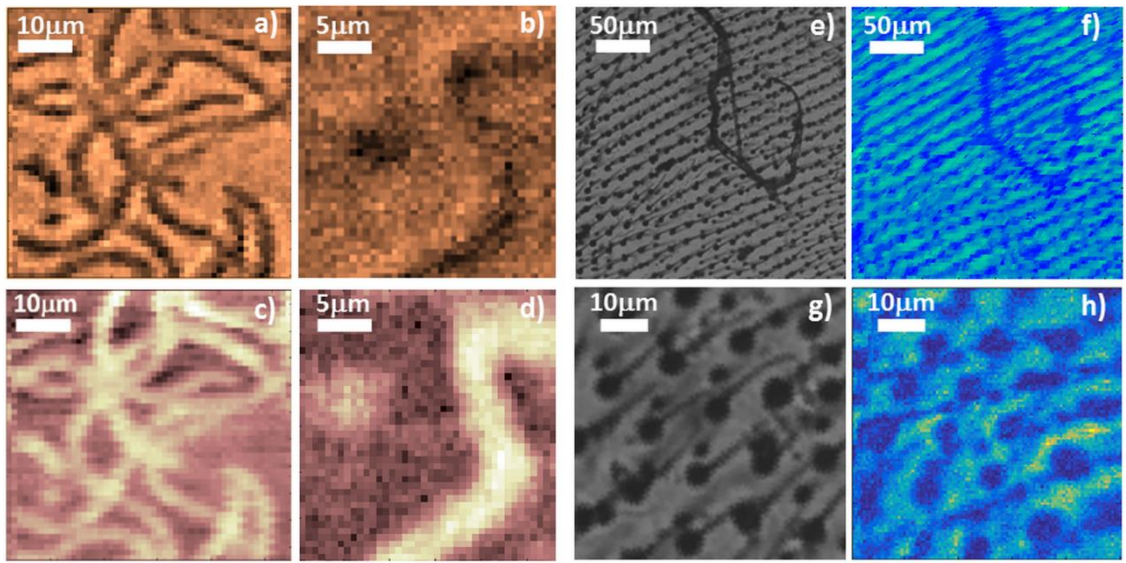

**Figure 4.** Amplitude (a and b) and phase (c and d) images of Spirulina filaments at different scan areas and step sizes (50 µm by 50 µm and 1 µm in a) and c), while 25 µm by 25 µm and 0.5 µm in b) and d), respectively). Normal microscopic (e and g) and SIMUP amplitude images (f and h) of a fruit fly wing at different scan areas and step sizes for the latter (250 µm by 250 µm and 5 µm, and 50 µm by 50 µm and 0.5 µm, respectively).

### III. DISCUSSION

The above examples allow a direct comparison with related approaches of UP imaging of biological objects [1, 4, 19, 20]. The experiments demonstrated an order-of-magnitude improvement in lateral resolution as compared to previously published results. The resolution is expected to be further enhanced by increasing the NA of the objective pair, which is readily allowed by the confocal arrangement. Although in its present form, the resolution of SIMUP technique is lower than the state-of-the-art multiphoton fluorescence techniques used in the short-wavelength infrared regime (SWIR, between 1000 and 2000 nm) [18], it may be sufficient for special applications in tissue imaging [13]. An additional important feature of the method is that, in addition to the amplitude image, it also yields a quantitative phase map of the sample, contrary, e.g., to the conventional ghost imaging techniques that are able to reveal only amplitude information [19, 20, 37]. (By adapting the principles of phase contrast or holographic microscopies, it is possible to retrieve phase information by ghost imaging, too, however, this extension demands a considerable increase of complexity of the experimental arrangement [38]-[41].)

From the practical point of view, our measuring system is a scanning holographic microscope with a feature of wavelength conversion. When scanning the sample point-by-point, we determine phase and amplitude from a single image recorded by the camera. Alternatively, one could do single-pixel observation (similarly to [37]), as well, instead of using the camera, but then one has to scan also the phase at each point (e.g., with a spatial light modulator (SLM)), which takes extra time to scanning.

On the other hand, application of optical scanning techniques (e.g., by a Nipkow-disk or an SLM) [33], [35] is expected to drastically speed up data acquisition, and lower the light dose per unit area of the sample, in the present arrangement, too. It should be noted here that in a recent work, Paterova et al. have presented an ingenious arrangement for layer-by-layer quantum imaging of reflective objects based on a Michaelson interferometer, with an opportunity of point-by-point imaging in all the spatial dimensions [36]. Although, they did not present images of biological samples, their 10-micrometer-range in-depth resolution could be useful in some biological applications, as well, especially, where reflected or back-scattered light is utilized for imaging.

To conclude, in this pilot study we introduced the concept of SIMUP imaging, demonstrated its technical feasibility and showed examples of applications in biological microscopy. Confocal scanning was combined with observing interference (and not with fluorescence), with possible implications in other types of scanning microscopies [42], [43]. So far, we applied the technique for 2D-imaging, but with proper modifications and modeling it can be extended to monitor 3D-objects, as well, similarly to [36]. Follow-up studies are going to clarify these points, and make SIMUP a powerful tool, with special applications in imaging objects whose extinction (either absorption or scattering or both) does not allow observation at one of the wavelengths (e.g. in the case of silicon chips [4, 20]). The most important applications, however, are envisioned in the investigation of sections of solid-state or biological samples that are absorbent or highly scattering in the visible range. The full "physical" (both transmission- and refractive index-wise) mapping of solid-state materials or biological tissues by the SIMUP technique may complement the results of "chemical" or "functional" imaging methods, such as CARS [44] or multiphoton fluorescence [17], [18], [35] microscopies.

### IV. MATERIALS AND METHODS
*Experimental setup*

The details of the experimental setup are shown in Fig. 5. The light of a cw Nd-YAG laser (LR1, Coherent, Verdi-V5 diode pumped Nd-YVO$_4$ laser, 532 nm) was splitted into two branches (BR1 and BR2) by a polarization beam splitter (PBS). The beam in BR2 was focussed into a ppKTP nonlinear crystal (type 0, 9.325 µm poling period, Raicol Crystals Ltd.), and the induced daughter beams, carrying the entangled, down-converted photon pairs of characteristic wavelengths of 880 nm and 1345 nm (spectral bandwidth of ca. 10 nm, each), were separated by a dichroic mirror (DM1). The 1345-nm light was then collimated by a lens (L2) and subsequently focussed onto the sample plane (S) by the objective, Obj1 ($f$ = 6.24 mm, NA = 0.4). The light transmitted through the sample was collected by the confocally placed Obj2 (an identical pair of Obj1), whereafter, via DM2, it was focussed into NL2, an identical pair of NL1, by L2'. The temperature of NL1 and NL2 was kept constant at 28°C by home-made aluminum sample holders equipped by Peltier thermostates, each. The 532-nm light from BR1 was also focussed into NL2 (by L1'), and generated another pair of 880-nm and 1345-nm daughter beams, the latter of which is indistinguishable from the light coming through the sample. A filter (F) then blocked the 532-nm and the 1345-nm beams, and transmitted only the 880-nm one, that was eventually hitting the camera, D (I-PENTAMAX-512-EFT/1, Princeton Instruments).

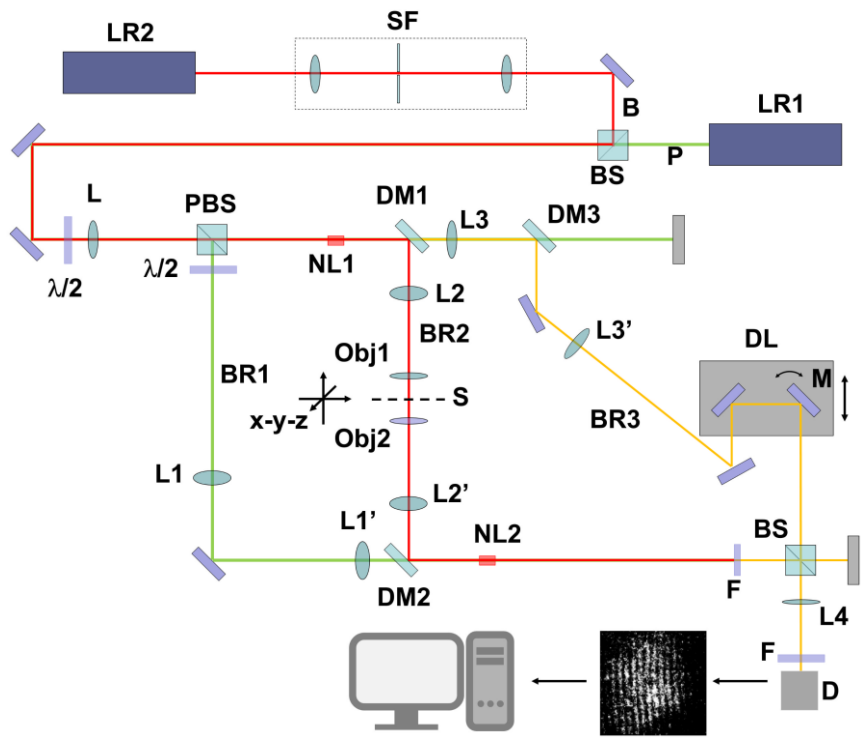

**Figure 5**. Scheme of the setup. The used abbreviations are as follows: LR1: CW green laser (532 nm) emitting the exciting beam, P; LR2: CW infrared laser (1342 nm) emitting the auxiliary background beam, B; DM1 and DM2: dicroic mirrors, reflecting the 1345-nm, while transmitting the 532-nm and 880-nm beams; NL: PPKTP crystal, type-0, splitting 532-nm photons into 880-nm and 1345-nm ones; PBS and λ/2 waveplate: polarization beam splitter cube, in combination with a waveplate to adjust the power distribution among the branches (BR1 versus BR2 and BR3); L: lens focusing the pumping beam into crystal NL1 ($f$ = 150 mm); L1 and L1', L2 and L2', L3 and L3': achromatic doublets, $f$ = 75 mm, L1 (532 nm), L1' (400 - 700 nm), L2 (1050 - 1700 nm), L3 (650 - 1050 nm); Obj1 and Obj2: confocally positioned achromatic dublet objectives (600 – 1050 nm), $f$ = 6.24 mm, NA = 0.4; BS: non-polarizing beam splitter cube; SF: spatial filter for OPA laser, focusing lens $f$ = 125 mm, 10-µm pinhole, $f$ = 60 mm collimator lens, incoming beam diameter 10 mm; L, L1, L2, L3: confocal arrangement; F: 3-nm band filter, adjusting the coherent length to ~ 150 µm (without the auxiliary laser); Delay line: mirrors and mechanical positioner to compensate up to 40 mm difference in optical pathlength. Mirror M was used to adjust the number of interference fringe lines on the camera.

The 880-nm light from NL1 was deflected by DM3 to the BR3 branch, and, after passing through a beam splitter (BS), also hit the camera, D. (All optomechanical parts were purchased from ThorLabs Inc., while the dichroic mirrors were manufactured by OPTILAB Ltd.) To reduce the effects of mechanical instability, the setup was mounted on a vibration isolated optical table, and it was covered by a home-made plastic hood, in order to avoid unwanted effects from air convection.

Using similar experimental arrangements, Zhou et al. [26] and Lemos et al. [1] showed that daughter photons of the same wavelength coming from NL2 are coherent with those travelling through BR3, due to induced coherence, if the optical pathlengh difference between light beams travelling from NL1 to D through BR2 and BR3 is within the coherence length of the system, that was estimated to be about 100 μm [1]. This condition was met by a delay line in BR3, adjusted by a mechanical positioner of 20-mm span by better than 5-μm precision. Note, that our setup contained an extra pair of objectives in BR2, as compared to the arrangement of Zeilinger et al. [1, 4], introducing an inevitable intensity loss in the sample path, accounted for by the factor η (see also in SI Appendix). In order to improve the signal-to-noise ratio of the images, an auxiliary solid-state laser (RLTMIL-1342-200, Changchun New Ind. Ltd.) was also used to provide weak background light of 1342 nm, matching the spectral band of one of the daughter beams coming from the nonlinear crystal (LR2 in Fig.5). In this case, the original signal photon number after NL1 ($n_1$) increased by a factor of nearly an order of magnitude ($1+n_B$), and by 1.5 after NL2 (see in SI Appendix), while the bandwidths of the signal modes decreased, due to the OPA effect [45]. Correspondingly, the coherence length of the detected signal beams considerably increased (to ca. 7 mm, according to our measurements).

By tilting mirror M around a vertical axis, interference fringes appearing as vertical stripes were generated on the screen of D. The number of stripes was adjusted to be around 10. (According to our experience, more than that did not improve resolution.) The data acquisition by the camera was executed with a frame rate between 1 and 5 fps, depending on the exposure time adjusted to the level of the measuring light. Considering the photon numbers per pixel, the number of pixels, the quantum efficiency of the camera, an upper limit of $3\cdot10^7$ and $10^8$ photons/s from the crystal was estimated for without and with seeding.

From the contrast of the interference images without object, one could estimate their maximal visibility, which was typically $0.62 \pm 0.05$ for our experiments. A series of grey filters (T = 0.3, 0.4, 0.6 and 0.8), were applied to determine the visibility versus transparency dependence, showing a moderate deviation from linearity. The experimental values were compared with the results of our quantum optical calculations (SI Appendix, Fig. S5), andthe $n_1$, $n_B$ and $η$ values were determined to be 0.44, 5.96 and 0.245, respectively. The images were stored on a computer, and their analysis was performed by a MATLAB program implementing an FFT routine, which yielded an amplitude and phase value to each image recorded.

During data acquisition, the samples were moved by a combination of computer-controlled translation scanners. A high-precision double-axis motorized mechanical stage (Scan IM Tango controller, Märzhäuser Wetzlar GmbH) and a 3-D piezo scanning stage (P3D 20-100, Spindler and Hoyer Inc.) were used for coarse and fine 2-D (X-Y) positioning, respectively. The Z-axis of the piezo scanner was utilized to adjust the sample to the common focal plane (S) of Obj1 and Obj2.

*Samples*

As a test object (T) to determine the resolution of the system, we used a grid of rectangular stripes on a glass substrate (No. 1.5 cover slip, 170 μm thickness), produced by photopolymerization using a direct laser writing system (μPG-101, Heidelberg Instruments GmbH). The specifications of the the grid were: EpoCore negative tone photoresist (Micro resist technology GmbH) of 1.5 um thickness, stripe-width 2 μm, grating constant 10 μm, both in the X and Y directions.

The spirulina strain was NIES-39, *Arthrospira platensis* Gomont. The cells were sandwiched between cover slips of 200-μm spacing, and fixed to the sample holder for the measurements.

The wing of a garden fruit fly was prepared and fixed to the sample holder as a free-standing object. During the measurements, the whole setup was covered by a plastic hood, in order to avoid air turbulence.

**APPENDIX**

See Supplemental Information in separate file.

**ACKNOWLEDGMENT**

The technical help of Zoltán Imre, László Dér, Dr. Anna Mathesz and Dr. Lóránd Kelemen is gratefully acknowledged. The work was supported by grants of the Hungarian National Research, Development and Innovation Office, "Photonic Applications of Biomaterials: NKFI-1 K124932" (for AB and AD), and "Development and applications of multimodal optical nanoscopy in material and life sciences: GINOP-2.3.2-15-2016-00036" (for MGB), the latter co-financed by the European Regional Development Fund. The spirulina strain was a kind gift of Dr. Péter Kós. Helpful discussions with Professor András Patkós on the theoretical principles are highly appreciated.

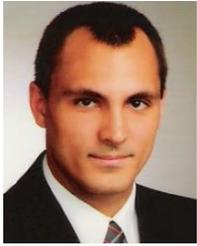

**ANDRÁS BÚZÁS** got his M.Sc. diploma from physics at the Szeged University in 2005. From 2005 to 2008, he was working as a Ph.D. student at the Department of Optics and Quantum Electronics, University of Szeged, Hungary.

In 2008, he joined the Bionanoscience Unit of the Institute of Biophysics, Biological Research Center of the Hungarian Academy of Sciences, and currently works there as a Research Associate in the Biophotonics and Biomicrofluidics research group.

His scientific interest includes diverse problems in biophotonics.

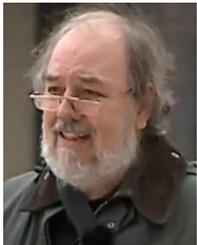

**ELMAR K. WOLFF** studied physics, and received his M.Sc. and Ph.D. degree at the Technical University of Dortmund, Germany.

He spent his post-doc years at the Department of Physics of the Massachusetts Institute of Technology (1974-1976) and at the Technical University of Dortmund (1976-1979), performing spinor experiments in Spin 1 Systems with NMR, and developing NMR multi frequency excitation techniques. From 1979 to 1984, he was a senior scientist in the Max Planck Institute for Ernahrungsphysiology, Dortmund, Germany. From 1984 to present, he works as the science director of the Insitute of Applied Biotechnology and System Analysis at the University of Witten-Herdecke, Germany.

He does research in Microbiology, Biotechnology and Bioinformatics. His current projects are bR application to all-optical logic and switching, Quantum Microscopy and strategies for analyzing complex systems via the newly proposed contextual system theory.

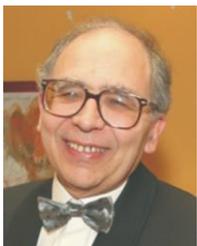

**MIHÁLY BENEDICT** is a professor emeritus of the Department of Theoretical Physics of the University of Szeged, Hungary.

His research interest includes fundamental problems of quantum theory from the point of view of their applications mainly in light-matter interactions on the atomic level and in solid state structures, as well. He is author or co-author of ca. 100 scientific papers in those fields, including the renowned monograph Superrradiance (IOP Bristol 1996).

He received the C.Sc. degree from the Leningrad (St.Petersburg) State University and holds the D.Sc. degree of the Hungarian Academy of Sciences. He served as the vice-president of the Roland Eötvös Physical Society (Hungary) 1996-99, and the representative of individual ordinary members in the EPS Council 2005-2009. He is also interested in the problems of physics education at the university level.

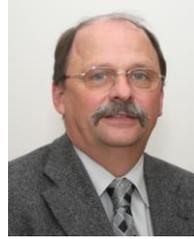

**PÁL ORMOS** had been the director general of the Biological Research Center of the Hungarian Academy of Sciences for 8 years.

He has M.Sc. (1975) and Ph.D. (1982) in physics from Jozsef Attila University Szeged. He is an ordinary member of the Hungarian Academy of Sciences (1994). P. Ormos is honorary president of the Hungarian Biophysical Society. Between 2002 and 2005 he served as vice president of IUPAP (International Union of Pure and Applied Physics).

P. Ormos's research is focused on single particle observation, optical micromanipulation and nanobiotechnology. He is author or co-author of ca. 100 publications and there are over 4000 citations to them. He is an elected Fellow of the American Physical Society.

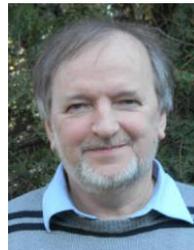

**ANDRÁS DÉR** is a scientific advisor of the Institute of Biophysics, Biological Research Center of the Hungarian Academy of Sciences, Szeged, Hungary.

He received an M.Sc. diploma in physics in 1980 at the József Attila University, Szeged, Hungary, a C.Sc. (PhD) and a D.Sc. degree from the Hungarian Academy of Sciences in 1988 and 1999, respectively. He had been the deputy director of the Institute of Biophysics, BRC, the co-president of the Biophysics Committee of the Hung. Acad. Sci., and the vice president of the Hungarian Biophysical Society for more than a decade.

His major research area is protein dynamics, biophotonics and bioelectronics. Together with his colleagues, he is named as inventor on several granted and pending patents about photonic applications of biomaterials.